\documentstyle[12pt]{article}                    
\newcommand{\beq}{\begin{equation}}             
\newcommand{\eeq}{\end{equation}}               
\newcommand{\bqry}{\begin{eqnarray}}            
\newcommand{\eqry}{\end{eqnarray}}              
\newcommand{\bqryn}{\begin{eqnarray*}}          
\newcommand{\eqryn}{\end{eqnarray*}}            
\newcommand{\preprint}[1]{\begin{table}[t]      
            \begin{flushright}                  
            \begin{large}{#1}\end{large}        
            \end{flushright}                    
            \end{table}}                        
\newcommand{\PD}[2]                             
    {\frac{\partial^{#2}}{\partial #1^{#2}}}    
\begin{document}
\preprint{LA-UR-98-3390}
\title{Comment on ``Regge Trajectories for All Flavors''}
\author{\\ L. Burakovsky\thanks{E-mail: BURAKOV@T5.LANL.GOV} \ and \ 
T. Goldman\thanks{E-mail: GOLDMAN@T5.LANL.GOV} \
\\  \\  Theoretical Division, MS B285 \\  Los Alamos National Laboratory \\ 
Los Alamos, NM 87545, USA }
\date{ }
\maketitle

\bigskip

In a recent Letter \cite{FPS} (and in ref. \cite{FS}), Filipponi, Pancheri and
Srivastava report on the construction of a formula for linear Regge 
trajectories for all quark flavors:
\beq
\alpha _{j\bar{i}}(t)=0.57-\frac{(m_i+m_j)}{{\rm GeV}}+\frac{0.9\;{\rm GeV}^2}{
1+0.2\;\!(\frac{m_i+m_j}{{\rm GeV}})^{3/2}}\;\!t,
\eeq
where $m_i,m_j$ are the corresponding constituent quark masses for the $j\bar{
i}$ trajectory. 

 
As the authors of \cite{FPS,FS} remark, no unique quark mass can be extracted
from (1), and each trajectory $\alpha _{j\bar{i}}(t)$ rather corresponds to 
its own set $(m_i,m_j).$ The values of $m_i$ can be extracted by using the
vector meson masses with hidden flavor into Eq. (1): $\alpha _{i\bar{i}}(M^2_{
i\bar{i}})=1.$ Such an extraction gives (in GeV, $n=u,d,$ and the superscript 
indicates the trajectory from which the corresponding value is extracted)
$m_n^\rho =0.05,\;m_s^\phi =0.23,\;m_c^{J/\psi }=1.70,\;m_b^\Upsilon =5.12.$
Then, the values of $m_i$'s for the $j\bar{i},\;i\neq j$ trajectories should 
be related to the above hidden-flavor values by additivity of trajectory 
intercepts. This additivity is satisfied in 
two-dimensional QCD and many QCD-motivated models
(\cite{BG} and references therein), 
and therefore should be considered as a firmly 
established theoretical constraint on Regge trajectories. 
It is easily seen that in the case of the trajectories (1), this
constraint implies 
$m_n^\rho +m_c^{J/\psi }=m_n^{D^\ast }+m_c^{D^\ast },$ 
$m_s^\phi +m_c^{J/\psi }=m_s^{D_s^\ast }+m_c^{D_s^\ast },$ 
$m_n^\rho +m_b^\Upsilon =m_n^{B^\ast }+m_b^{B^\ast },$ 
$m_s^\phi +m_b^\Upsilon =m_s^{B_s^\ast }+m_b^{B_s^\ast }.$
Thus, e.g., the parameters $m_i$ of the $D^\ast $ and $D_s^\ast $ trajectories
must be related to those of the $\rho ,\phi $ and $J/\psi $ ones, even if no 
unique values of $m_i$ can be extracted. Using now these parameters as given 
by the above relations for calculating the vector meson masses, through 
$\alpha _{j\bar{i}}(M^2_{j\bar{i}})=1,$ one finds (in MeV) $M(D^\ast )=
1882.5,$ $M(D_s^\ast )=2007.1,$ $M(B^\ast )=4566.3,$ $M(B_s^\ast )=4724.1,$
in contrast to the measured values \cite{pdg} (in MeV) $M(D^\ast )=2008\pm 2,$
$M(D_s^\ast )=2112.4\pm 0.7,$ $M(B^\ast )=5324.8\pm 1.8,$ $M(B_s^\ast )=5416.3
\pm 3.3.$ In the last two cases, the discrepancy between the calculated and 
measured values is $\sim 700$ MeV which is an unsatisfactorily large 
inaccuracy. Thus, the trajectories (1) cannot combine both meson spectroscopy 
and additivity of intercepts; fixing the parameters $m_i$ to reproduce 
spectroscopy will necessarily result in violation of the intercept additivity
constraint. We note that simple constituent quark model relations, e.g., 
$M(B^\ast )=(M(\rho )+M(\Upsilon ))/2,$ $M(B_s^\ast )=(M(\phi )+M(\Upsilon ))
/2,$ give better values than Eq. (1): (in MeV) 
$M(B^\ast )=5114,$ $M(B_s^\ast )=5240.$  
Moreover, the numerical values of intercepts given by (1) in the light quark 
sector contradict data. Indeed, Eq. (1) gives $\alpha _\rho (0)=0.47,$ vs. 
$\alpha _\rho (0)=0.55,$ as extracted by Donnachie and Landshoff 
from the analysis of $pp$ and $p\bar{p}$ scattering data \cite{DL}, and 
$\alpha _{K^\ast }(0)=0.29,$ vs. $\alpha _{K^\ast }(0)\approx 0.40$ as follows
from the analysis of hypercharge exchange processes $\pi ^{+}p\rightarrow 
K^{+}\Sigma ^{+}$ and $K^{-}p\rightarrow \pi ^{-}\Sigma ^{+}$ \cite{VKKT}. 
Since the values of intercepts determine the $s$-dependence of the total 
cross-sections, $\sigma _{tot}\propto s^{\alpha (0)-1},$ and the differential 
cross-section profiles, $d\sigma/dx_F\propto (1-x_F)^{1-2\alpha (0)},$ it is 
among the requirements for the theory to predict the exact numerical values of
intercepts.   



In ref. \cite{FS}, two of the authors notice that since the flavor dependent
Regge slope $\alpha ^{'}=\alpha ^{'}(0)/(1+A\tilde{m}),\;\tilde{m}=m_i+m_j$ 
has a large negative derivative for small $\tilde{m},$ it appears that the 
condition on all the slopes in the light quark sector $\alpha ^{'}\sim 0.8-0.9$
GeV$^{-2}$ can be satisfied only with almost exact mass degeneracy in this 
sector. This fact, as noticed in ref. \cite{FS}, prevented the authors from
constructing trajectories satisfying additivity of inverse slopes which is
another constraint provided by the heavy quark limit \cite{BG}, in addition to
intercept additivity, which the trajectories (1) do not meet. Although
their remark is correct, we disagree that $\alpha ^{'}=\alpha ^{'}(0)/(1+A
\tilde{m})$ is the only form that may be used in order to construct the 
trajectory. Indeed, as we discuss in \cite{BG}, the form
$\alpha ^{'}_{j\bar{i}}=\frac{4}{\pi }\;\!\frac{\alpha ^{'}}{1+\sqrt{
\alpha ^{'}}\;\!(m_i+m_j)/2},$
where $\alpha ^{'}=0.88$ GeV$^{-2}$ is the standard Regge slope in the light
quark sector, satisfies additivity of inverse slopes, and reproduces the 
values of the slopes in agreement with those extracted from data, for the 
following constituent quark masses (in GeV):
$m_n=0.29,\;\;m_s=0.46,\;\;m_c=1.65,\;\;m_b=4.80,$
which, in contrast to the above values given by (1), are not 
atypical of values used in phenomenological quark models.

We believe this analysis raises serious doubts as to the suitability of the 
formula (1) for the phenomenological description of quarkonia.

\end{document}